\documentstyle[12pt,aasms4]{article}

\lefthead{Keel et al.}
\righthead{Large-Scale Structure at $z \approx 2.4$}

\begin{document}

\title{Evidence for Large-Scale Structure at $z \approx 2.4$ From 
Lyman $\alpha$ Imaging}

\author{William C. Keel\altaffilmark{1}}
\affil{Department of Physics and Astronomy, University of Alabama,
Box 870324, Tuscaloosa, AL 35487}
\authoremail{keel@bildad.astr.ua.edu}

\author{Seth H. Cohen\altaffilmark{1}, Rogier A. Windhorst\altaffilmark{1}, and 
Ian Waddington\altaffilmark{1}}
\affil{Department of Physics and Astronomy, Arizona State University,
Box 1504, Tempe, AZ 85287}

\altaffiltext{1}{Visiting astronomer, Kitt Peak National Observatory, 
National Optical Astronomy Observatories, operated by AURA, Inc., under 
cooperative agreement with the National Science Foundation.}


\begin{abstract}
The history of large-scale structure depends on cosmological parameters 
and on how merging unfolds among both galaxies and groups. There has been
recent evidence for clustering among Lyman-$\alpha$ emitters, Lyman-break
galaxies, and Lyman absorbers. We present deep wide-field medium-band imaging in
redshifted Lyman $\alpha$ of fields surrounding regions selected to have
{\it HST} detections of faint Lyman $\alpha$ emitters, over a range
of surface densities, to characterize the larger-scale environment.
The radio galaxy 53W002 was previously found using {\it HST} to be part of a 
rich grouping at $z=2.39$, including $\approx 5$ spectroscopically confirmed,
compact, lower-luminosity star-forming 
objects. Our new data show this to be part of a larger structure traced by 
bright active nuclei, all contained within a projected span of 6.8\arcmin\  
(3.2 Mpc). 
Of the 14 candidate emitters, six have been spectroscopically confirmed as 
active nuclei in the range $z=2.390 \pm 0.008$. Various statistical tests
give a significance of 95-99\% for the reality of this structure on the sky.
Our data thus strengthen the evidence for clustering at these redshifts.

The grouping around 53W002 is more extended than a relaxed King model, at the
90\% confidence level. This may be evidence either for a configuration
which has yet to decouple fully from the Hubble expansion, or for multiple
subgroupings which will themselves at some point form a more compact,
relaxed structure. The redshift range for measured members is comparable to the
Hubble flow across the structure, which may imply that the structure is
seen near turnaround and suggests that its mass cannot be derived from
the velocity dispersion.

We surveyed two additional 14\arcmin\  fields at $z=2.4$, 
each centered on an HST WFPC2
field which has been searched for faint Lyman $\alpha$ emitters, as
well as three contiguous fields near 53W002 for objects at $z\sim2.55$. Only a
single emitter consistent with showing Lyman $\alpha$ at $z \approx 2.4$ 
appears in these fields, to a somewhat brighter flux limit, while a total
of six candidate emitters appear in the three fields at $z \sim 2.6$. 
Comparison with the (deeper but narrower) WFPC2 surveys suggests that the 
surface-density contrast from field to field is larger for brighter objects.
From this survey alone, groupings such as the
53W002 ``cluster" must have an area covering fraction $< 0.04$ in this redshift
range.

Three of the AGN in the structure at $z=2.4$ show extended emission-line
structure, with equivalent widths suggesting {\it in situ} photoionization
as far as 50 kpc from the nuclei.
Such objects may populate deep surveys as diffuse
Lyman $\alpha$ emission clouds when their cores are sufficiently obscured. 
\end{abstract}

\section{Introduction}

As our ability to measure galaxy evolution has grown, so has the possibility
of observing how galaxy clustering has evolved. On various linear scales,
this is relevant to the merger rate of both galaxies (and perhaps their
smaller progenitors) and groups, and to cosmological parameters. As
reviewed recently by Cen (1998), the growth of representative cluster
masses depends strongly on $\Omega_0$. This is due not to a direct relation
between $\Omega_0$ and structure growth {\it per se}, but more to the fact 
that the calculations must be normalized to match the present-epoch mass 
spectrum, which introduces a coupling between the amplitude
$\sigma_8$ of the power spectrum and $\Omega_0$ for viable models. 

Several studies have indeed shown evidence for cluster-scale structures at 
redshifts $z > 2$. Our {\it HST} imaging (Pascarelle et al. 1996b, herefter 
P96b; Pascarelle et al. 1998, hereafter P98), using a combination of broadband 
and medium-band filters to isolate Lyman $\alpha$ emission in the relevant 
redshift range, showed that the $z=2.4$ radio galaxy 53W002
is part of a rich assemblage of Lyman $\alpha$ emitters. Most of these are
compact (effective radius $r_e \approx 0.1$" or 0.8 kpc), and are powered by
star formation rather than by classical active nuclei. In P98, we showed that
the surface density of such objects varies between different random 
lines of sight by approximately a factor of 4, with the 53W002 field being the 
richest we have observed thus far. A somewhat different grouping at similar 
redshift ($z=2.38$) was identified by Francis et al. (1996, 1997), who
found four Lyman $\alpha$ emitters very close to the redshifts of Lyman
$\alpha$ absorbers seen against two background QSOs. These emitters are seen 
over a projected span of 0.63 Mpc, and are much redder than the objects found
by P96b. In an analogous way, Malkan et al. (1995, 1996) used narrow-band 
near-infrared imagery to find three H$\alpha$--emitting objects at $z=2.50$ in 
the foreground of the QSO SBS 0953+545 at $z=2.58$, closely matching the redshifts 
of metal-line absorption systems seen in the QSO spectrum. Starting from 
a sample of Lyman-break galaxies, Steidel et al. (1998, also Steidel 1999) 
have found 
a concentration of galaxies at $z=3.090 \pm 0.015$ spanning about
$4 \times 8$ Mpc. And at even higher redshift, Hu \& McMahon (1996) report 
spectroscopically-confirmed Lyman $\alpha$ companions to the $z=4.55$ QSO 
BR2237--0607.

These results show that it is now possible to trace developing clusters,
and other large-scale structure, at high redshift. The new generation of 
wide-field imagers has enabled survey strategies that
can tell how common, how extensive, and of what amplitude structures
are in the galaxy distribution at various redshifts. As a first step in this
direction, we present here a Lyman $\alpha$ survey of large fields around
the regions we have searched with {\it HST}, to place the object counts
from those fields in a larger context, and in particular to probe the 
spatial extent and bright end of the luminosity function of the
cluster which includes 53W002.

In evaluating size and luminosity, we use $\rm{H_0}=80$ km s$^{-1}$ Mpc$^{-1}$,
$q_0=1/2$, which gives an angular scale of 128" per Mpc. Scaling for other 
values, linear sizes scale directly with $\rm{H_0}$ and luminosities
as $\rm{H_0}^2$. For other values of $q_0$, as a shortcut, we note that 
linear sizes (luminosities) quoted here would be multiplied by 2.9 (8.2)
for $q_0=0.1$ and by 0.49 (0.24) for $q_0=1$.

\section{Observations}

We observed several fields around the radio galaxy 53W002 at $z=2.39$ (Windhorst
et al. 1991), which had been shown from imaging in redshifted Lyman $\alpha$
to be part of a structure containing additional AGN and star-forming 
galaxies (P96ab, P98). This field therefore offered
a unique opportunity to probe a known structure at significant redshift.
We observed two further WFPC2 fields
using the same filter set, as part of a parallel survey for additional
objects in the window around $z=2.4$ (P98). For comparison, we also
observed a large region adjacent to the 53W002 area using a filter tuned
for Lyman $\alpha$ emission at $z \sim 2.55$.

For the current wide-field extension of the {\it HST} medium-band survey, we
used the PFCCD imager, with $2048^2$ Tektronix CCD, for observing
runs in the 1997 and 1998 summer seasons on the 4m Mayall
telescope of Kitt Peak National Observatory. At the time, this system
had significantly better throughput at 4100--4300 \AA\  than the wider-field
Mosaic system. Each exposure covered a region 14.3\arcmin\  on a side 
with 0.420\arcsec\  pixels. We isolated Lyman $\alpha$ in the redshift ranges
$z=2.32-2.45$ and $z=2.49-2.61$ with intermediate-band filters, the first of which
was intended as a clone of the WFPC2 F410M filter, manufactured by Custom 
Scientific, Inc.,  to the same specifications as the HST filter set. We refer 
to these as F413M and F433M to avoid confusion with the WFPC2 F410M filter; 
WFPC2 has no close counterpart to F433M. These filters have FWHM=150 \AA\  and 
peak transmission at 4150 and 4330 \AA\  respectively, as measured in a parallel 
beam; the peak transmission moves blueward by $\sim 12$ \AA\  and the FWHM 
increases by about $\sim 19$ \AA\  in the $f/2.7$ prime-focus beam of the
Mayall telescope (Marcus 1998). In addition to the medium-band Lyman $\alpha$ 
filters, we also observed each field in $B$, for continuum magnitudes, and 
$V$ to account for color terms in the continuum subtraction (as in 
P96b). As it happened, we were able to observe three contiguous fields
just to the northeast of 53W002 in F433M. No Lyman $\alpha$
emission candidates in the overlapping region were common to both filters,
except for the brighter of the new QSOs discussed later, whose Lyman $\alpha$
emission is so strong that it was detected even in the extreme wings
of the redder filter's passband.

Total exposure times for each region and filter are listed in Table 1, with the 
area of full exposure extending 420\arcsec\  in each coordinate from the listed
position. Individual exposures were 30 minutes for the medium passbands, and 
10-20 minutes in the broad bands, with dither motions of 20-30 arcseconds 
between successive exposures to suppress residual flat-field and cosmetic
effects in the stacked combination images. The various field
pointings were sometimes shifted from an exact rectangular pattern
to avoid stray light from bright stars within a region extending
about 5\arcmin\  outward from the CCD edge. The image stacks show image FWHM in the
range 1.2--1.6\arcsec\ . The number of objects detected in each field depends
on both seeing and total exposure times, and the number shown in Table
1 reflects detections in both $B$ and medium-band filters. The $B$ limiting
magnitude is given for each field using a $3 \sigma$ threshold.

%add seeing FWHM to table 1

%                  F413M      B         V
%        ngc6251  31.3052   32.7852   32.9879
%        w02      31.3052   32.9150   32.6200
%        huar     31.2365   32.8627   33.0201

The broadband data were converted to standard Johnson $BV$
magnitudes via secondary standard stars in M92, NGC 7006, and NGC 4147 
(Christian et al. 1985, Odewahn et al. 1992). The photometric zero points
were consistent to 0.02 magnitude or better from night to night. Both
$B$ and $V$ magnitudes show color terms at the 0.03-magnitude
level per unit change in $(B-V)$. A more important issue is that 
of the color correction in continuum subtraction, as outlined below.

The 1997 data suffered from an additive ghost image of the telescope
pupil occupying much of the field, produced by internal reflections in 
the optical corrector. This ghost image was not present in the 1998 data, since 
the dewar had been offset from the optical axis to avoid the problem.
As an additive artifact, the ghost image could be isolated by comparing
medium- and broad-band sky flats, then removed by 
subtracting scaled versions to eliminate the
ghosting in our stacked images as completely as possible. Many of the images
suffered from spatially variable background structure in the ``blank sky" 
regions due
to scattered starlight from stars both within and outside the field
of view, sometimes modulated by passing cirrus clouds, which we subtracted
using a $101 \times 101$-pixel (42") median filter, clipped around
the brightest galaxies which would otherwise be partially subtracted. 
This allowed higher quality in the final average images, since pixels would 
not be artificially flagged for rejection because of a temporarily high 
background. This leaves spurious residual dark halos around bright stars, but
since these are additive, local background subtraction will still give
accurate photometry quite close to such stars.

We identified emission- and absorption-line candidates starting with 
object lists and photometry generated using version
1.0a of SExtractor (Bertin \& Arnouts 1996),
using visual inspection to reject putative detections which were
compromised by bright stars or artifacts near the edges
of individual exposures comprising the stacked mosaics. The detection
parameters were: object detection threshold 2.5$\sigma$ above background 
over 5 contiguous pixels, and a deblending parameter 0.005 (which turned
out to be essentially irrelevant at this level of crowding). Table 1 
includes numbers of objects in each field appearing in the matched
$B,V, m_{413}/m_{430}$ catalogs. Detections in all three bands were
required to deal with color terms in the continuum-to-line comparison.
The relative exposure depths suggest that we should not be missing
comparable objects due to color effects (though the possibility of
extreme colors, such as very red objects, still exists).
This multiband matching requirement means that the listed detection totals 
do not simply reflect the relative exposure times. Coordinates were
measured by fitting a celestial coordinate system to stars from the
HST Guide-Star Catalog (GSC) on each frame; the formal accuracy is
0.25" rms, borne out by recovering positions of individual GSC stars.

The threshold for emission-line detection is not completely straightforward, 
since each object's detectability depends both on the line flux and its 
equivalent width. Our primary criterion was for equivalent width incorporating 
individual error estimates, with a secondary list using formal significance of 
line emission as a basis for selection. Since the F413M medium-band filter sits 
on the blue edge of the $B$ passband, there is a color term accounting for the 
continuum slope between $B$ and 4130 \AA\  (a similar but smaller
term exists for the F433M filter). We follow W91 and P96B in using the traced 
filter properties to compute the locus of featureless power-law spectra (a 
reasonable approximation for galaxies in the emitted ultraviolet) in the 
$(m_{413}-B)-(B-V)$ plane, as shown in Fig. 1 for the 53W002 field. 
This locus is well approximated 
by the line
$$ (F413-B) = 0.32 (B-V) - 0.08, $$
which is a good fit to the observed distribution of ``field" objects in
our data (the numerical constants become 0.10 for the slope and 0.0 for the
intercept in the case of the F433M filter). 
Our primary sample of emission-line candidates consists of
objects which fall more than $4 \sigma$ below this relation where
$\sigma$ applies to the scatter of points on the emission side of 
the distribution's ridge line (Fig. 1), which puts our threshold
at 0.6 magnitude
in F413M/F433M excess (observed equivalent width about 110 \AA\ , corresponding 
to an emitted equivalent width 30-32 \AA\  at $z=2.4-2.6$). These candidates are
listed in Tables 2 and 3 for the two filters, and enlargements of the 
intermediate-band and $B$
images are shown in Figs. 2 and 3. Here, the tabulated $(m_{413}-B)$ and 
$(m_{430}-B)$ have been corrected for first-order
color terms as described below; negative values indicate an excess in the
narrower passband. The listed equivalent widths are in the
observed frame; the emitted value will be smaller by $ (1+z) \approx 3.4-3.6$.
The Lyman $\alpha$ EW and flux for the three objects previously reported by 
P96b -- their object numbers 18 and 19 plus 53W002 itself -- are somewhat 
uncertain because
each has a resolved Lyman $\alpha$ emission region (see section 6 below), 
which produces somewhat
different values depending on how the flux is extracted. There is evidence
that object 19 is itself variable as well (P96a).

The reliability of this 
sample is supported by the fact that all the members observed spectroscopically
(in the 53W002 field) are indeed active nuclei at $z=2.39$. To assess
whether there is an additional population of detections with lower
equivalent width but comparable statistical reliability, we also considered
object selection by significance in the deep 53W002 F413M data, defined as
$$ S = {{[(F413 - B) - 0.32 (B-V) +0.08 ]}/ \sigma}$$
(where $\sigma$ here is the statistical error in the F413-B color)
with the additional requirements of $B > 23.5$ and computed
equivalent width $\ge$ 90 \AA\  to avoid
spurious detections of bright objects where the formal errors are
much smaller than the scatter introduced by spectral features in
stars and lower-redshift galaxies. All of the detections in the primary
list have significance $> 5 \sigma$ by this criterion. Within the
range of $B$ and $S$ that contains all the primary detections,
the 53W002 field includes an additional five candidates, thus potentially
augmenting the total number by about one third (which occur all over the field,
unlike the clustered equivalent-width candidates). These are listed
at the bottom of Table 2, but since this technique doesn't generate any
additional objects with line flux significantly above the threshold
of the original list (even while relaxing the possible error bounds), we 
concentrate on the equivalent-width defined list. This list includes
some objects with line fluxes as low as 
$3.7 \times 10^{-17}$ erg cm$^{-2}$ s$^{-1}$, but a characteristic flux
limit for approximate comparison with other results would be close to
$5 \times 10^{-17}$ for the 53W002 field in the F410M filter. Corresponding 
values for the other fields at are $1.4 \times 10^{-16}$ for HU Aqr
and $1.0 \times 10^{-16}$ for NGC 6251 and the three fields observed 
with the F430M filter.

Contamination of the Lyman $\alpha$ sample by objects with [O II] $\lambda 3727$
emission at $z \simeq 0.11$ (F413M) or $z=0.15$ (F433M)
should not be important, for the following reasons. The Keck spectroscopy
of several of our candidates, plus objects from the {\it HST} lists, by
Armus et al. (1999), shows that all the emission-line objects either
have multiple lines at $z \sim 2.4$ or a single line with equivalent
width plus continuum shape inconsistent with [O II] as judged from nearby
objects. Finally, these objects are all smaller than 1\arcsec\  in effective radius
(that is, the blue continuum is either unresolved or almost so from the
ground) and fainter than $B=23$, which would translate simultaneously 
into linear extent of less than 4 kpc and absolute magnitude fainter than 
$M_B=-15$ for objects at redshift low enough to have [O II] emission in our  
passbands.

% significance candidates' x,y coordinates are
% 621, 865      958, 1403     1561, 686    471, 1847    1267,256

% EW assuming filter effective width 150 A
% F(Ly alpha) in cgs units, assuming centered in the passband

These data are not deep enough to recover the star-forming objects seen
by Lyman $\alpha$ emission in the WFPC2 data of P96b and P98, even in the 
deepest Kitt Peak
F413 exposure on the 53W002 field. The brightest such candidate in the
HU Aqr WFPC2 field from P98 also falls slightly below our equivalent-width 
limit. These known $z=2.4$ objects
have emission-line intensities which would correspond to Kitt Peak detection
levels typically $0.5 \sigma$, which is confirmed by comparison of our 
ground-based detections. Therefore we are
tracing structures using objects which are, as far as we can tell from
the spectroscopically identified subset, fairly luminous active nuclei.
For 53W002 and its two immediate neighbors, {\it HST\/} imagery shows these to be 
accompanied by dimmer star-forming objects. Whether these are smoothly
distributed throughout the clumping or form smaller structures in the
regions traced by AGN is an important question for further work.

The two strongest-emission new candidates in the 53W002 field, bright 
enough to appear
on blink inspection of the data during the observing run, were observed 
spectroscopically using the Ritchey-Chretien spectrograph and T2KB CCD at the
Mayall telescope a week after the 1997 imaging observations.
Grating KPC-10A (316 lines/mm, first-order blaze at 4000 \AA\ )
gave 2.77 \AA\  pixels with usable sensitivity over the
3700--8000 \AA\  region, and resolution typically 2.1 pixels = 5.8 \AA\ 
FWHM. Each object was acquired by offset from bright stars, as measured on the
CCD frames. Mediocre seeing mandated a relatively wide 1.7\arcsec\  slit opening,
and even so the seeing was variable enough that only single 60-minute
exposures of high quality were obtained for each object. While conditions
were not photometric, the data could be placed on a relative flux scale using 
observations of the hot standard star PG 1708+602 (and the objects'
broadband magnitudes are accurately known from the multiband imagery).

Both new candidates observed were found to be QSOs, as shown in Fig. 4.
Redshifts were measured by taking centroids of the bright emission lines,
by Gaussian fits to the profile peaks, and by cross-correlation with the
mean QSO spectrum assembled by Francis et al. (1991), adopting a mean
of these redshift measures and using the differences as a measure of the error.
Their spectroscopic
properties are given in Table 4, with redshifts measured from
Lyman $\alpha$ and C IV individually and from cross-correlation. Both
have redshifts within $\Delta z = 0.005$ of the other objects in the
53W002 grouping (P96ab, Armus et al. 1999). We also include emitted-frame
Lyman $\alpha$ widths, to make the point that these are fairly
narrow-lined QSOs.

As this paper was in revision, we received word that Pascarelle, Yahil, \& 
Puetter (1999) have confirmed candidate 6 from Table 2 with a redshift
close to $z=2.38$ based on Keck spectroscopy, giving a total of six 
confirmations of the imaging candidates.

%\begin{figure}%[tbd]
%\plotone{highzfig4.eps}
%\plotfiddle{highzfig4.eps}{400pt}{-90}{60}{60}{-250}{350}{0}
%\caption{}
%\end{figure}

\section {Emission-Line Detections and Field-to-Field Variations}

Our most striking result is seen from Tables 2 and 3, and Fig. 5, where we
show the distribution of candidate Lyman $\alpha$ emitters in part of the
53W002 F413M field. 
There is an extensive grouping of emission-line candidates including
53W002 which has no counterpart in the other fields observed at
either $z=2.4$ or $z=2.55$. Fourteen objects in the 53W002 field passed our 
equivalent-width criterion for line emission, while
only one at the edge of the HU Aqr field did, and none in the NGC 6251
field. Among the F413M fields, with differing exposure times,
the field-to-field ratios to a highest common limiting line flux would be 4:1:0 
(the first value rising to 6 for the intermediate limit appropriate to the 
empty NGC 6251 field). There are 6 candidates in the $z=2.55$ range
sampled by the F433M filter, over almost three times the solid angle
of the 53W002 field (and twelve times the solid angle encompassing the
candidate emitters in that field).
This shows that there 
are significant structures in place at $z=2.4$, but not necessarily
over a large fraction of the sky; a simple estimate based on these
data alone is that such assemblages cover less than 0.04 of the sky
in a redshift range $\Delta z = 0.15$, with the limit arising from the
fact that the 53W002 field was observed precisely because we already knew
that some additional objects were present. The amplitude we find from 
field to field is even greater
than that found by P98 from fainter HST detections at the center of each
of these fields, which might indicate that luminous AGN are more clumped
than fainter objects. This could mean, for example, that the more
massive objects (thus more likely to host AGN)
start life more strongly biased toward initial mass peaks.

We can assess the statistical significance of the grouping around 53W002
in several ways. First, we address the reality of the clumping seen within the
53W002 field at $z=2.4$. Most simply, the probability of $n$ objects falling
within a single region covering a fraction $f$ of the solid angle
surveyed will be $p = f^{(n-1)}$, where the location of the region
is not otherwise specified. In this case, using the superscribed circle
about the 14 candidates for the region size and the area of full exposure
in the stacked F410M as the overall field surveyed, $f=0.38$ and 
$p= 4 \times 10^{-6}$. A Monte Carlo simulation indicates a somewhat
higher (though still small) probability of having 14 points drawn
from a uniform random distribution fall within a circle of this size,
$p = 0.0014$. Finally, we used the two-dimensional version of the
Kolmogorov-Smirnov test as proposed by Peacock (1983) and examined
in detail by Fasano \& Francheschini (1987), employing the routines
presented by Press et al. (1992). This test gives a significance level
of 95\% for the clustering within the 53W002 field. The critical values for
the 2-dimensional test are slightly dependent on the distribution of sample
points, but these points are not strongly correlated (Pearson $r=0.40$)
and the effect would only act at the $\pm 1$\% level in this regime.
This most conservative of the tests still shows the grouping with high
significance.

To test the sigificance of variations in the number of objects from field to
field, we use combinatorics to ask how likely it is that a uniform
distribution sampled with our total number of detections (to a common
flux limit) would be so strongly weighted toward a specified field
(since we already knew from previous data that there is an excess around 
53W002). As noted above, to the highest of the three limiting fluxes,
there are four objects in the 53W002 field, one in the HU Aqr field, and
none in the NGC 6251 parallel field. Of the 207 ways to distribute
5 objects among three bins, 11 are at least this strongly weighted to
the specified one, yielding a probability of $11/207=0.053$ of achieving
this result by chance. Thus the significance of the higher number of
objects in the 53W002 field is 95\%, even without taking into account
their concentration within the observed area in this field. We note that
there is somewhat weaker evidence for clumping of the detections in
the F430M filter ($z \sim 2.55$) in the 53W002 NE field; these objects all
have derived line luminosities in the QSO range.

\section{Lyman-$\alpha$ absorption candidates}

Many of the brightest Lyman-break galaxies observed by Steidel et al. 
(1996, 1998) show net absorption at Lyman $\alpha$, indeed sometimes with no
significant emission. One of the factors contributing to the
strength or weakness of the emission may be metallicity, through the enhanced
formation of grains which can absorb resonantly scattered Lyman $\alpha$
photons (Bonilha et al. 1979), though observations of Lyman $\alpha$
in starbursts of different metallicity show that there must
be more to the story than this single parameter (Giavalisco et al. 1996,
Lequeux et al. 1995, Thuan \& Izotov 1997). Using imaging techniques, we are 
sensitive only to {\it net} emission, while some of the line emission may be 
cancelled by the line in absorption from stellar atmospheres and H I in the 
galaxy. Indeed, about 1/3 of the Lyman-break galaxies observed by Steidel et al. 
(1998) show net absorption at Lyman $\alpha$. Therefore, we consider here the 
possibility of detecting objects with strong absorption at Lyman $\alpha$.

As a guide to the strength of Lyman $\alpha$ absorption expected from
star-forming galaxies, we use the HST GHRS spectrum of the bright knot
in NGC 4214 obtained by Leitherer et al. (1996). After excising the narrow
emission component, this spectrum shows an absorption line of equivalent
width $\approx 28$ \AA\ , which would be an observed value of EW=95 \AA\ 
at $z=2.4$. This is just within our detection threshold for objects to
$B=25$ in the 53W002 field, so that we can extract absorption candidates
in the same way as the emission candidates. Since even luminous star-forming
galaxies are unlikely to exceed QSO luminosities, we restrict the
selection to objects in the range $24<B<25$, thereby avoiding much of
the potential confusion from foreground stars and galaxies which have
an absorption edge between the continuum and narrowband filters, and using 
the equivalent-width criterion to screen out stars with strong H$\delta$
absorption. 
This is a particular issue for white dwarfs,
which would also be distinguished by broadband colors much bluer than
expected for any high-redshift galaxies. Accordingly, we restrict the
candidate absorption objects to the range $0.15 < (B-V) < 1$ and require
significance of the absorption to exceed $4 \sigma$. This color range
is wider than we observe for the star-forming emitters in the WFPC2
data (P96b). These criteria leave 4 candidates in the 53W002 field
(Table 5, Fig. 6). Three of these are in same spatial region as the emission
candidates, but our ability to select these objects in a more
precise way is limited by the fact that the scatter in the $(m_{413}-B), (B-V)$
diagram is asymmetric and larger to the absorption side, largely due to the
natural signal-to-noise limitations at faint levels, so that
there are many more interlopers at a given equivalent width for absorption
than for emission.

%\begin{figure}%[tbd]
%\plotone{highzfig6.eps}
%\caption{}
%\end{figure}

\section{The 53W002 ``Cluster" at $z=2.39$}

These results strengthen the evidence for some sort of clustering at
early cosmic times. We consider here what kind of assemblage we
see in the 53W002 field, and how it might relate to the clustering we
see today. This entails measures of its size, population, and dynamical state.

\subsection{Cluster ``Size" and Radial Distribution}

The virial radius 
$R_v = {{1}\over {n}} [\Sigma_{j < i} 
({{1} \over {{ \vert \bf r}_i - {\bf r}_j}\vert })]^{-1}$
of this assemblage of 14 objects is 157" or 1.2 Mpc in proper coordinates,
which would correspond to
1.9 Mpc (a factor $\pi/2$ larger) in three dimensions for a typical projection
geometry. The radial distribution is so extended that fewer than half the
candidates (four) lie within this projected radius of the centroid.

For a distribution this sparsely sampled, whose centroid is
not well determined by a strong central concentration, it may be more enlightening
to consider the fraction of objects encompassed by circumscribed
(projected) circles than by such a specific physical
measure as the virial radius. 
All fourteen candidates are contained within a
radius of 327", with 2/3 (10) contained within r=218" and half
(seven) inside r=164".

To examine whether the distribution of these objects looks like contemporary
relaxed systems, we consider how well a King-law profile with 
any core radius can be
fitted to our observations. Because the center is ill-defined from such
a sparse sample, we use as a statistic for comparison the cumulative
number of objects within an encircled radius, whose center can
drift to accomodate the maximum number within a given 
radius. From $10^4$ Monte Carlo samples of 14 objects each drawn from a King profile
(in number density), we generated the 
bounds containing various fractions
of the trials and identify these with confidence intervals. 
Since the $z=2.4$ structure is traced by active nuclei
at B$\le$ 24.5 -- which may occur in rather
low-luminosity galaxies this close to the peak redshift for QSO number
density -- we use a model for number density rather than incorporating 
some level of mass segregation to represent the luminosity density in a 
typical rich cluster.
The core radius is also determined by fitting the radial scale of the
distributions so that the true significance of each band may be slightly
greater. The core radius was left as an adjustable parameter to be
determined by the best fit to the observed cumulative distribution.
This exercise should tell whether the observed two-dimensional distribution 
is likely to be drawn from one like the relaxed profiles of nearby (rich)
clusters, and if so what its radial scale (the core radius for a King model) is.
As shown in Fig. 7, the 53W002 association is less centrally condensed than
a King model of any core radius. Specifically, if either the inner or
outer four points in the number-radius relation are used to anchor the
data to the Monte Carlo predictions, some points fall outside the 90\%
band (and if the inner points are fit, outside the 97\% band). 
The difference is such that the inner points imply a core radius of
76" (0.6 Mpc), while the outer ones a core radius of 42" (0.3 Mpc).
At this 90-97\% confidence level, we can reject a relaxed King distribution 
for these objects. 

If this grouping is not yet relaxed, we are left with an ambiguity in 
intepreting its linear scale -- would it evolve more nearly in comoving
or proper coordinates? If it has yet to turn around from the Hubble
expansion, it will grow for some time in proper coordinates
but shrink (slightly) in comoving ones until it turns around.
On the other hand, it may have already turned around but not yet
have virialized, in which case the proper-coordinate linear scale will
remain nearly constant. Other reports of high-redshift structures
have made differing assumptions on this matter -- note that Steidel et al.
(1998) quote a comoving extent for their structure at $z=3.1$, while
some other workers use proper length. For the 53W002 structure, some 
intermediate case would be most appropriate, still allowing a wide range of
current length scales for comparison with the present epoch.

We can address the correlation function $w(\theta)$ as measured in the
53W002 field from these data. While it is clearly a high-amplitude structure, 
this measurement might furnish
useful information on length scales as well as just how large an 
amplitude could be reached by $z=2.4$. 
Edge effects were assessed by Monte Carlo
trials, employing the expression from Landy \& Szalay (1993) as used by
Neuschaefer \& Windhorst (1995). At $\theta=30$\arcsec\ (200 kpc), 
$w(\theta )=3.2$, 
and we see positive 
correlation out to a radius $r=5.5$\arcmin\ (2.4 Mpc) above a threshold
value $w \sim 1$. Following the treatment by Neuschafer \& Windhorst (1995),
and using their sample to $g=25$, thus result does indicate that the
region around 53W002 (sliced in both angle and redshift)
is more strongly clustered than the field by a factor $\sim 3$, since the
``field" objects have an amplitude of only about 0.03 covering a redshift
range roughly $z=0.5-2$.

\subsection{Galaxy and AGN Content}

The objects from our emission candidate list which have been spectroscopically
confirmed are all obvious AGN, with a mix of broad- and narrow-lined cases.
This is not surprising given the flux constraints on spectroscopy with
4-m class telescopes, but
it is already an unusual AGN population for a single group. Their
absolute magnitudes are in the range associated with, for example, 
low-redshift PG quasars ($M_B=-21.4$ to $-22.4$ for our adopted cosmology,
following Weedman 1986 in dealing with spectral slope).
The brightest objects known from the WFPC2 field {\it not} to be
such AGN have $M_B \sim -20.5$, so it remains unclear what the fainter
KPNO detections at $B=24-25$ represent. Certainly these would not be
unusual luminosities for additional AGN, but this is a regime in
which star-forming objects are not unreasonable either.

One hint might come from image structure. If AGN are weaker for the fainter
objects, they might appear more clearly resolved since the core is less
dominant. We compared image FWHM values (from the SExtractor tables and
as computed by the IRAF {\it imexamine} task) for the emission candidates
with those of bright, unsaturated stellar images nearby in the $B$ frame.
Except for the extended structures around 53W002 and object 18, which
have substantial line-emission components, all the candidates are
unresolved. This fits with the sizes of the objects in the WFPC2 field,
whose typical half-light radii are 0.10\arcsec\ , but doesn't furnish any further
constraints on whether these new objects are more likely to be AGN or
bright star-forming systems.

Even for only passive evolution of the star-forming objects, it is significant
that we see none brighter than $M_B=-21$, consistent with the {\it HST}
results but now covering a much larger region. A typical $L^*$ galaxy would
have $M_B=-23$ at $z=2.4$ unless either active evolution continued
to substantially lower redshift, or merging of these small objects
continued to form today's luminous galaxies. More luminous galaxies
could hide by lacking Lyman $\alpha$ emission, perhaps if they are
more metal-rich and hence can suppress emission in this line, or if their
star formation rate dropped quickly at early times. Near-IR line
surveys could test the first possibility.
%and we find no such objects...
 
\subsection{Velocity Dispersion}

The five spectroscopically confirmed members of the 53W002 grouping
have a velocity dispersion $\sigma_z=0.0060$, translating into 
$\sigma_v=532$ km s$^{-1}$ in the objects' frame. Adding the two
additional faint emitters from P98  drops this value to 467 km s$^{-1}$.
While one should respect the errors in estimating the velocity dispersion
from such small samples, it does seem clear that we are not dealing with
the dynamics of a rich virialized cluster with $\sigma_v = 1000$ km s$^{-1}$.

% QSOs: 2.381, 2.393; W02   2.390; 18    2.393; 19    2.397

Since many of the previously reported {\it HST} objects around 53W002 are 
apparently
star-forming complexes, with very narrow Lyman $\alpha$, there is the
possibility of introducing a systematic offset in comparison
with redshift measurements of broad-lined AGN using the same line.
This is less of an issue with the three narrow-lined (``type 2") AGN
previously reported in this region (P96a,b). Comparison of the strong UV
lines (Lyman $\alpha$, C IV, C III]) with lower-ionization species or
with narrow emitted-optical
lines expected to arise far from the core (especially [O II] $\lambda 3727$)
have shown that substantial differences in central velocity can exist.
The shifts can exceed 1000 km s$^{-1}$ for radio-loud objects, but have
a mean close to zero for radio-quiet QSOs (Espey et al. 1989,
Marziani et al. 1996).
In addition to being radio-quiet (Richards et al. 1999), the two 
new QSOs have rather narrow lines 
compared to many of the
ones studied for velocity shifts (and compared to the Francis et al. 1991
composite), so the shifts may not be as large. In fact, their close match
to the redshifts of other objects in the field would be a remarkable
coincidence if systematic shifts of more than a few hundred km s$^{-1}$ are
present, but the possibility remains that the actual velocity range of these
objects is larger than the value we measure from Lyman $\alpha$ and C IV
alone.

Furthermore, since the radial distribution suggests that the structure is
not virialized and may still be coupled to the Hubble flow, we consider the 
limiting case in which the velocity range represents the Hubble flow across 
the depth of the structure rather than internal motions driven by gravity.
For $q_0 = 1/2$ (or its $\Omega+\Lambda$ counterpart), the Hubble 
parameter $H$ would have
been greater at $z=2.4$ than today's $H_0$ by a factor about
6 (scaling inversely
with cosmic time for this cosmology), so that
the relevant expansion rate would have been in the range 300--600
km s$^{-1}$ Mpc$^{-1}$ for ${\rm H}_0=$50--100 km s$^{-1}$ Mpc$^{-1}$. 
For a characteristic line-of-sight depth of 1.5 Mpc, 
comparable to the observed transverse extent containing most of the members, 
this implies a ``velocity dispersion" of 450-900 km s$^{-1}$ even for
a completely unbound assemblage. Since the positional data show clearly that 
the grouping has decoupled from the Hubble flow to the extent of
showing a density contrast of at least a factor 4, we interpret this
comparison as showing that this group is still turning around from the
Hubble expansion, so that the velocity data do not necessarily allow
us to measure its mass (or anything else about the detailed dynamics).

\section{Lyman $\alpha$ Haloes of Constituent AGN}

Three of the bright AGN in this field show extended
Lyman $\alpha$ structures in WFPC2 data (P96b, P98). These are
either linear or roughly biconical, fitting with a general paradigm
of ionizing radiation directed mostly along the poles of some disklike
structure. The KPNO data have better sensitivity to large regions of low
surface brightness than does {\it HST}, and reveal new aspects of the 
extended line emission.
For 53W002 and object 19 (in the P96b nomenclature), this is an extension of 
the Lyman $\alpha$ structure seen in WFPC2 images (Windhorst, Keel, \& 
Pascarelle 1998), but for object 18, this resolved structure is not
only much larger (extending more than 5\arcsec\  from the core) than the 
ionization or scattering cone inferred from
WFPC2 data, but it is most extended in a different direction.
The inner parts of these structures are detected as well in H$\alpha$
using IRTF narrowband imagery and in [O III] using NICMOS multiband and grism
data (Keel et al. 1999).

We can examine the structure of the Lyman $\alpha$ images by comparing both
the $B$ continuum and emission-line images of each candidate emitter to
stellar profiles from the same region of each image. This gives some 
insurance against minor PSF changes across the field, and avoids problems 
due to somewhat different PSF widths between the $B$ and F413M images. 
We consider extended emission to be detected when there is some scaling
between broad-- and medium--band images for which the difference is flat
across the core and shows flux more extensive than the PSF.
Requiring a flat central profile is conservative, to minimize the possibility 
of false detections at the expense of underestimating the flux in the
spatially extended component. Several of the brightest emission-line objects 
show Lyman $\alpha$ emission more extended than their continuum structures.

This analysis suggests that both scattering and local recombination play roles
in these emission-line halos. The three objects with extended emission-line
regions illustrate this:

\noindent Object 18 (P96b): The PSF subtraction shows that more than 75\% of the 
Lyman $\alpha$ 
flux from object 18 comes from outside the core, and recovers the gross 
features of the WFPC2 image. Similar results come from analysis of the
$B$ image, while the relative count rates indicate that most of the
$B$ light is in fact Lyman $\alpha$. Spectroscopy by Armus et al. (1999)
shows that the extended cloud has almost no continuum component,
consistent with these results. This accounts for the very blue color of
the extended structure ($(B-V)=-0.4$), since there are no strong emission
lines in the $V$ band.

\noindent 53W002: For 53W002 and object 19, about half the line flux is 
spatially resolved, in accord with the HST PC data of Windhorst et al. (1998) and 
the ground-based Lyman $\alpha$ imaging from Windhorst et al. (1991). As 
noted earlier,
the emission-line structure is approximately along the orientation of
the 1\arcsec\  radio double source, but much larger. 

\noindent Object 19 (P96b): As in 53W002, about half the line flux is 
resolved, in accord with the {\it HST} data as well. In this case, the extended
emission is all in Lyman $\alpha$ to our detection threshold; less than
10\% of the $B$-band flux comes from outside the core.
 
These extended structures are illustrated in Fig. 8, comparing the
medium-band image, the PSF-subtracted version, and HST imagery of the
brightest regions. The large-scale line emission is well aligned
with the small-scale emission observed with {\it HST}, which is
well shown in the color figure of Windhorst et al. (1998) including
scattered continuum components. For object 18, the KPNO data reveal
that the inner emission region is identical with the two major components
seen with {\it HST}, but much more extensive and amorphous material
appears at this deeper surface-brightness threshold.

For the two newly detected QSOs,
any such resolved line-emitting region must have less than 10\% of the 
total Lyman $\alpha$
flux (and as low as 5\% for the brighter QSO 2). These values apply to
structures that are extended on the scale resolved by the PFCCD images;
as a guide, the image size in the final F413M stack has 1.2\arcsec\  FWHM.

Lyman $\alpha$ emission by itself is difficult to interpret, since we
lack useful density indicators and its radiative transfer is sensitive
to the velocity field and dust content. At a minimum, if mechanical energy
input isn't important in the extended nebulae, the number of Lyman $\alpha$
photons can give a lower limit to the number of ionizing photons reaching
the gas, provided only that the situation is in a steady state.
In turn, this can tell us whether the radiation field must be anisotropic
to account for the structures we see - that is, whether we are correct in
referring to some of these structures as ionization ``cones". The continua
on our line of sight are measured from about 1100-2000 \AA\  in the emitted 
frame, so that we should be able to do a reasonable extrapolation to the
Lyman limit and estimate the expected number of ionizing photons in the
isotropic case. For the simple case of a photoionized cloud occupying
solid angle $\Omega$ as seen from the central source, if we see the
same ionizing continuum as the cloud does, the extrapolated continuum
and observed Lyman $\alpha$ emission should satisfy
$$ n_{LyC} \ge {{\Omega} \over {4 \pi}} {{n_{Ly \alpha}} \over {f_\alpha}} $$
where $n_{LyC}$ is the number of Lyman continuum photons per second 
extrapolated from the observed continuum, $n_{Ly \alpha}$ is the observed number 
of Lyman $\alpha$ photons per second, and $f_\alpha$ is the fraction of
recombinations whose cascade includes Lyman $\alpha$ (0.64 for 
case A, following the tabulations in Osterbrock 1989). The luminosity
distance has cancelled on both sides, though we still need to make a plausible
assumption about the clouds' geometry to assign a subtended $\Omega$. The 
equality
holds for an ionization-bounded nebula which is optically thin to all the
Lyman lines, in the sense that violating these conditions increases the
continuum/line ratio and therefore makes the observed continuum more
sufficient to power the extended line region.

% critical global EW for crossover is about 900 A * (omega/4 pi)

Applying this test to the three resolved Lyman $\alpha$ regions shows that
at least object 18, with its very extensive line emission, has an ionization
source that we don't see. Extrapolating the observed continuum at its
flat level in flux falls short of creating the observed Lyman $\alpha$
emission by at least a factor 2, suggesting either a bump in the
ionizing spectrum or anisotropic radiation. The $B-K$ continuum shape
is not unusually red, in fact quite normal for narrow-line AGN and
almost identical to the other two objects with extended line emission, so
that anisotropic illumination makes sense if it is not caused by material
that would redden the observed continuum. 
A similar issue appears for many type 2 Seyfert nuclei, with a Lyman-continuum
deficit implied by the observed continuum and line intensities, and blue
UV continuum slopes. This has been variously attributed to scattering or 
reflection of radiation from a small continuum region (as in
Antonucci, Hurt, \& Miller 1994), and surrounding star formation (Colina et al. 
1997), with recent results suggesting that the nucleus itself may not 
be an important contributor to the UV flux in narrow-line objects.
The geometry of the Lyman $\alpha$ cloud near
object 18 offers little help; while the inner parts, as detected with
{\it HST} (P96a, Windhorst et al. 1998) resemble an ionization cone, 
the outer regions
are extended at $90^\circ$ in projection to this axis. Of course,
additional energy sources might be considered, such as the radio jet
interactions proposed for powerful radio galaxies. However, of these
three objects, only 53W002 itself has significant resolved radio emission;
the other two are both substantially weaker and unresolved by the VLA
at the 1\arcsec\  level (Richards et al. 1999). The flux data alone do not 
require anisotropic radiation
for 53W002 and object 19, though the emission-line structure at least suggests
an anisotropic gas distribution, and it is suspicious that the 
Lyman $\alpha$ structure in 53W002 aligns with the smaller double radio
source (Windhorst, Keel, \& Pascarelle 1998).

Independent of the ionization mechanism, such a rich collection of large
clouds around the brightest illuminating sources raises the question
of whether the extended gas belongs exclusively to the AGN hosts or
exists more widely throughout this cluster, where we cannot observe it
so easily. Detection of these structures in the continuum at a level above
the weak free-free emission accompanying recombination would imply
the presence of dust, a tracer of the level of star formation early
in the galaxies' history. Furthermore, if the nuclei are more often
obscured early in cosmic time, we might expect to see ``disembodied"
Lyman $\alpha$ clouds in upcoming deep surveys. 

\section{The Evolution of Structure: Forward to the Past}

We have reported a Lyman $\alpha$ survey aimed at tracing structures
in the range $z=2.3-2.6$, finding a clumping or clustering in one
field, represented by 14 luminous objects spanning about 3 Mpc. This
adds to the existing evidence for structure in place, if not
necessarily well developed, at cosmologically early epochs.

What does the 53W002 structure turn into? Based on its extent as 
found here, we can ask how many members might exist to the {\it HST}
detection threshold. The existing WFPC2 data cover only a single 
5.7-arcminute$^2$ area, while we find candidate members spread over an area
of about 93 arcminutes$^2$. If the WFPC2 field is representative, there
would be 16 times as many faint star-forming members as we've detected
to date. With 8 objects in the {\it HST} field now spectroscopically 
confirmed as members (Armus et al. 1999), that means the total membership
would surpass 120 if we're seeing a smooth distribution. Alternatively,
if the star-forming objects are in clumps traced by the AGN that we
detect from KPNO, there would still be $\sim 70$ in this structure.
These numbers are lower estimates, since objects undoubtedly
occur below our detection thresholds. These two cases represent rather
different proposed histories for cluster and group formation -- in one
case, that clumps of objects will merge into today's galaxies, and
in the other, that individual objects we see at $z=2.4$ will either
passively evolve as they begin to exhaust their gas or continue to acquire 
infalling material, with merging of initially separate galaxies a less
important process.

Our velocity information is largely confined to the original WFPC2
field, with the addition of the two newly-identified QSOs. Thus
it is not very clear how the small velocity dispersion of these
objects (of order 385 km s$^{-1}$, including redshifts of members from
Armus et al. 1999) should be interpreted for the whole structure.
Furthermore, the extended spatial distribution suggests that the structure
has not yet relaxed, and may not yet be fully decoupled from the 
Hubble flow. Recent simulations of galaxy formation from a clumpy
medium by Haehnelt, Steinmetz, \& Rauch (1998) indicate that line-of-sight
velocity measurements not only have a factor 2 dispersion as seen from
various directions, but underestimate the relaxed virial velocities
by $\sim 60$\%. 
These considerations all make a virial mass estimate
very uncertain, and likely a lower limit. It may be more realistic 
to consider the velocity dispersion as applying to the megaparsec-scale
clumping including 53W002 itself and the AGN in objects 18 and 19. 
The velocity range we see is comparable to the dispersion expected
purely from the Hubble flow on an assemblage 3 Mpc deep at this epoch,
so we may well be seeing the group near the time of turnaround from
cosmological expansion, in which case the velocity dispersion tells
very little about the internal dynamics.

These questions suggest several potentially fruitful lines for further
work. Most notably, we need to know more about the content of this
structure, especially for fainter objects both with and without strong
Lyman $\alpha$ emission. Multiband imagery sufficient to derive photometric
redshifts and narrowband near-infrared measurements tailored to find
emission from [O II], [O III], or H$\alpha$ can help fill out our
census of members. A more accurate accounting of how common such structures
are in the early Universe will require wider-field multiband surveys,
preferably with fine enough wavelength bands to both pick out line
emitters and resolve multiple line-of-sight sheets or clusters.
Eventually, dynamical studies should tell us how these early assemblages
become the rich structural spectrum seen in today's Universe.

\acknowledgments{We are grateful to Richard Green for approving, and the 
KPNO staff for
implementing, a scheduling switch between the imaging and spectroscopic
observing runs which allowed us to confirm the new QSO candidates.
We acknowledge C. Leitherer and colleagues for making their starburst
spectral templates available via WWW. Portions of this work were supported by
NSF grant AST-9802963 and HST STScI grants GO-5985.0*.96A and
AR-8388.0*.98A. We thank Paul Francis, the referee, for goading us into
more quantitave probability assessments than we had originally incorporated,
as well as some interesting suggestions on cosmology versus dynamics in the
cluster redshift distribution.}

                                \clearpage
                                \title{
Figure Captions
                                }

%1

                               \figcaption{
Color-color selection of emisison-line candidates from the 53W002
field, showing the color term in the locus of most objects in the
two-color $(m_{413}-B),(B-V)$ plane. The dashed lines indicate the
$\pm 3 \sigma$ bands for emission and absorption candidates for our
equivalent-width selection. Emission candidates fall below the lower dashed line,
with some rejected because of neighbor contamination or nearby cosmetic
defects. Objects which we have spectroscopically confirmed are
marked with triangles.
                                }

%2

                               \figcaption{
Montage of F413M and $B$ subimages of the primary Lyman $\alpha$
emission candidates, as listed in Table 2. Each section is 31 pixels = 
13.0\arcsec\ 
square, with north at the top, with the candidate centered. The intensity 
scaling for all images in each 
filter is the same, so that different objects may be directly compared.
The relative scaling between F413M and $B$ images has been set so that
objects with no excess and neutral colors have approximately equal
intensity in both.
For compactness, each is labelled with the running number from Table 2.
The black area adjacent to candidate 16 in the B image is a charge-transfer
artifact from a bright star to the north. Candidates 1--6 have been 
spectroscopically confirmed at $z=2.39$. The final one, number 20, is in the
HU Aqr parallel field.
                                 }
%3

                               \figcaption{
Emission candidates at $z \sim 2.55$ from the F433M filter. 
As in Fig. 2, the medium-band and $B$ images of 13.0\arcsec\  regions are shown
centered on each candidate. The running numbers are as given in Table 3.
The area just southwest of object 25 is affected by a stellar reflection in
the medium-band interference filter.
                                 }
%4

                               \figcaption{
Spectra of the newly identified QSOs. The data for the brighter
object, candidate 2 = 171444.72+501744.9 from Table 2, have been shifted
upward by $5 \times 10^{-18}$ for clarity. Expected locations of the
weaker Si IV+O IV] and C III] features are marked for $z=2.393$.
                                 }
%5

                               \figcaption{
A portion of the 53W002 field as seen in the F413M filter, with
emission-line candidates marked along with the region searched by P96
using a similar filter on HST with WFPC2. This region covers 0.45 by area
of the entire
field observed, and contains all the Lyman $\alpha$ candidates. The two
additional marked objects within the WFPC2 footprint are numbers 
1 and 3 in our candidate list, and are spectroscopically confirmed
as AGN as shown by P96. The two newly discovered QSOs are marked as such.
                                 }
%6

                               \figcaption{
Absorption-line candidates at $z \sim 2.4$ from the F413M filter in the 53W002
%field, displayed as in Fig. 2. Again, each box is 13\arcsec\  square, with north at
the top. A pure continuum object of neutral color would have equal brightness
in both frames; color terms will distort this slightly for some of the
candidate absorbers.
                                 }
%7

                               \figcaption{
Minimum encircling radii for various numbers of objects among the
53W002 candidates versus Monte Carlo predictions for a King model.
The $\pm 2,3 \sigma$ bands are actually the 90 and 95th percentile
locations from the numerical trials. The data are plotted as normalized
to core radii of 42 and 76 arcseconds, which fit the inner and outer
portions, to illustrate that no single normalization fits without
falling outside the 90\% bound somewhere.
                                 }
%8

                               \figcaption{
Comparison of resolved structures in Lyman $\alpha$, with the observed
KPNO medium-band image on top, the PSF-subtracted version in the middle,
and the {\it HST} data from Windhorst, Pascarelle, \& Keel (1998)
in the bottom row. Each section is 10\arcsec\  square, with north at the top. 
The white triangular regions in two of the HST images shows where the
``gutter" between the WF2 and WF3 chips falls. The WFPC2 data for 
objects 18 and 19 in the P96 nomenclature are interleaved images taken
in a $2 \times 2$ dither pattern on 0.05\arcsec\  centers, with that
effective pixel size. For 53W002, the F410M
image is from the PC (Windhorst et al. 1998) and has
been smoothed with a Gaussian kernel of FWHM=0.23\arcsec\  to show the
faint resolved emission at about PA=290$^\circ$ from the nucleus.
The PSF is almost perfectly circular, so that the elongation of 
53W002 and object 19 at low levels in the KPNO data indicates resolved features.
                                 }

\noindent
\begin{tabular}{lcccrrrr}
\multicolumn{8}{c} {Table 1}\\
\multicolumn{8}{c}{Summary of Fields Observed}\\

          &  & \multispan2 Center (2000) & \multispan2 Exposure (min) & Matched & \\
Field    & Filter & $\alpha$ & $\delta$ &  F413/430M & B &    Objects &  B(lim) \\
53W002    &  413M &   17 14 10.3 & +50 16 07  & 480 & 100 & 3161  &   26.5 \\
HU Aqr par &  413M &  21 07 27.7 & -05 23 07  & 150 &  60 &  928  &   25.2 \\
NGC 6251 par & 413M & 16 36 37.0 & +82 34 10  & 180 &  75 &  930  &   25.5 \\
53W002 E  &   430M &  17 15 22.4 & +50 17 23  & 240 &  30 & 1588  &   25.0 \\
53W002 N  &   430M &  17 14 01.6 & +50 28 42  & 240 &  30 & 2215  &   25.5 \\
53W002 NE &   430M &  17 15 06.6 & +50 28 23  & 240 &  40 & 2167  &   25.5 \\                                         
\end{tabular}

%\clearpage
\noindent
\begin{tabular}{rccrrrrl}
\multicolumn{8}{c} {Table 2}\\
\multicolumn{8}{c}{Candidate Lyman $\alpha$ emitters from F413M, $z \approx 2.4$}\\

%No. & $\alpha$(2000) & $\delta$(2000) &  $B$ & $m_{4130}-B$ & EW (\AA\ ) & 
%F(Ly $\alpha$) & Notes \\
No. & $\alpha$(2000) & $\delta$(2000) &  $B$ & $(m_{4130}-B)$ & Obs. EW &
F(Ly $\alpha$) & Notes \\
    &                &                & (mag)& (mag)        & EW      &
(cgs) & \\

\multicolumn{8}{l} {53W002 field: primary equivalent-width sample} \\
 1& 17:14:12.01 & +50:16:02.3   & 23.41 & $-1.29 \pm 0.04$ & 342 & 1.1(-15) & P96b object 18 \\
 2& 17:14:44.72 & +50:17:44.9   & 23.31 & $-1.24 \pm 0.03$ & 320 & 1.1(-15) & New QSO \\
 3& 17:14:11.30 & +50:16:09.4   & 23.91 & $-1.01 \pm 0.04$ & 230 & 4.5(-16) & P96b object 19 \\
 4& 17:14:12.39 & +50:18:18.0   & 24.37 & $-1.00 \pm 0.04$ & 226 & 2.9(-16) & New QSO \\
 5& 17:14:14.70 & +50:15:29.7   & 24.05 & $-0.80 \pm 0.05$ & 164 & 2.8(-16) & 53W002 \\
 6& 17:14:39.82 & +50:21:52.3   & 25.46 & $-0.85 \pm 0.09$ & 178 & 8.4(-17) & \\
 7& 17:14:32.80 & +50:15:50.7   & 26.19 & $-0.78 \pm 0.15$ & 158 & 3.8(-17) & \\
 8& 17:14:31.66 & +50:19:06.8   & 25.54 & $-0.66 \pm 0.11$ & 125 & 5.5(-17) & \\
 9& 17:14:24.76 & +50:20:45.7   & 25.36 & $-0.62 \pm 0.09$ & 115 & 5.9(-17) & \\
10& 17:14:42.15 & +50:16:51.8   & 24.34 & $-0.73 \pm 0.08$ & 144 & 1.9(-16) & \\
11& 17:14:39.20 & +50:21:32.8   & 25.66 & $-0.83 \pm 0.13$ & 172 & 6.7(-17) & \\
12& 17:14:53.75 & +50:22:31.8   & 25.87 & $-0.62 \pm 0.15$ & 116 & 3.7(-17) & \\
13& 17:14:42.48 & +50:16:01.6   & 25.09 & $-0.67 \pm 0.09$ & 128 & 8.5(-17) & \\
14& 17:14:30.77 & +50:12:09.1   & 26.21 & $-0.82 \pm 0.16$ & 169 & 4.0(-17) & \\ 
            &               &  \\
\multicolumn{8}{l} {53W002 field: additional significance-selected candidates} \\
15& 17:13:49.35 & +50:14:29.0   & 25.05 & $-0.52 \pm 0.07$ &  92 & 6.3(-17) & \\
16& 17:14:04.15 & +50:18:14.6   & 25.24 & $-0.53 \pm 0.09$ &  94 & 5.4(-17) & \\ 
17& 17:14:30.46 & +50:13:13.7   & 25.58 & $-0.58 \pm 0.10$ & 106 & 4.4(-17) & \\
18& 17:13:42.86 & +50:21:20.4   & 25.35 & $-0.57 \pm 0.11$ & 104 & 5.4(-17) & \\
19& 17:14:17.55 & +50:10:13.6   & 25.50 & $-0.55 \pm 0.12$ &  98 & 4.4(-17) & \\
            &               &  \\ 
\multicolumn{8}{l} {HU Aqr field} \\
20& 21:07:57.26 & -05:25:41.1   & 22.25 & $-1.04 \pm 0.03$ & 241 & 2.2(-15) & \\
\end{tabular}

\clearpage
\noindent
\begin{tabular}{rccrrrrl}
\multicolumn{8}{c}{Table 3}\\
\multicolumn{8}{c}{Candidate Lyman $\alpha$ emitters from F433M, $z \approx 2.6$}\\
No. & $\alpha$(2000) & $\delta$(2000) &  $B$ & $(m_{4300}-B)$ & Obs. EW &
F(Ly $\alpha$) & Notes \\
    &                &                & (mag)& (mag)        & EW      &
(cgs) & \\
\multicolumn{8}{l} {53W002 N field:} \\
21& 17:13:17.36 & +50:27:12.8   & 24.55 & $-1.17 \pm 0.15$ & 291 & 3.2(-16) & \\
\multicolumn{8}{l} {53W002 E field:} \\
22& 17:15:23.23 &  +50:19:35.2  & 22.38 & $-1.17 \pm 0.05$ & 155 & 1.2(-15) & \\
\multicolumn{8}{l} {53W002 NE field:} \\
23& 17:14:51.62 &  +50:23:13.4  & 24.54 & $-0.87 \pm 0.16$ & 184 & 2.0(-16) & \\
24& 17:15:33.80 &  +50:28:49.6  & 24.56 & $-0.99 \pm 0.15$ & 223 & 2.4(-16) & \\
25& 17:15:28.13 &  +50:23:46.2  & 24.57 & $-1.02 \pm 0.16$ & 234 & 2.5(-16) & \\
26& 17:15:32.92 &  +50:30:52.1  & 24.25 & $-1.06 \pm 0.11$ & 248 & 3.5(-16) & \\

\end{tabular} 

\clearpage
\begin{tabular}{lrrccc}
\multicolumn{6}{c} {Table 4}\\
\multicolumn{6}{c}{Lyman $\alpha$ Absorption Candidates ($z\sim 2.4$) in the 53W002 Field}\\
No. & $\alpha_{2000}$ & $\delta_{2000}$ & $B$ & $(m_{413}-B)$ & $(B-V)$\\
27 & 17:13:44.51 & +50:21:08.9 & $24.01 \pm 0.04$ & $1.94 \pm 0.19$ & $0.18 \pm 0.10$\\
28 & 17:14:05.79 & +50:19:17.8 & $24.81 \pm 0.05$ & $1.46 \pm 0.18$ & $0.24 \pm 0.14$\\
29 & 17:14:23.37 & +50:23:09.2 & $24.25 \pm 0.04$ & $1.56 \pm 0.14$ & $0.57 \pm 0.07$\\
30 & 17:14:20.42 & +50:10:51.2 & $24.29 \pm 0.03$ & $1.56 \pm 0.14$ & $0.86 \pm 0.05$\\
\end{tabular}

\clearpage
\noindent
\begin{tabular}{lccccc}
\multicolumn{6}{c} {Table 5 \hfil}\\
\multicolumn6{c}{New QSO redshift measurements \hfil}\\
Object & Ly $\alpha$ FWHM (\AA\ ) &  Ly $\alpha$ & C IV & 
Cross-correlation & Mean \\
       &                          & \multispan2 (centroids) & \\
2 (171444.72+501744.9) & 11 & 2.384 & 2.377 & 2.382 & $2.381 \pm 0.002$ \\
4 (171412.39+501818.0) & 17 & 2.391 & 2.400 & 2.388 & $2.393 \pm 0.006$ \\                   \\
% QSO 2 was raw number 33; QSO 4 was number 36
\end{tabular}

\end{document}